\documentclass[12pt]{article}
\usepackage{amsfonts}
\usepackage{amsmath}
\usepackage{epsfig}

\begin{document}

\begin{flushright}
6/2008\\
\end{flushright}
\vspace{20mm}
\begin{center}
\large {\bf Further Development of the Tetron Model}\\
\mbox{ }\\
\normalsize
\vskip3cm
{\bf Bodo Lampe} \\              
e-mail: Lampe.Bodo@web.de \\   
\vspace{1.5cm}
{\bf Abstract}\\
\end{center}
After a prologue which clarifies some issues left open in 
my last paper, the main features of the tetron 
model of elementary particles are discussed in the 
light of recent developments, in particular 
the formation of strong and electroweak vector bosons
and a microscopic understanding of how the observed 
tetrahedral symmetry of the fermion spectrum may arise.

\newpage

\section{Prologue} 

In the left-right symmetric standard model with gauge group 
$U(1)_{B-L}\times SU(3)_c \times SU(2)_L \times SU(2)_R$ \cite{su2su2} 
there are 24 left-handed and 24 right-handed fermion fields 
which including antiparticles amounts to 96 degrees of freedom, 
i.e. this model has right handed neutrinos as well as 
righthanded weak interactions. 

In a recent paper \cite{lampe2} a new ordering scheme for the observed 
spectrum of quarks and leptons was presented, which relies on the 
structure of the group of permutations $S_4$ of four objects, and a 
mechanism was proposed, how 'germs' of the Standard Model 
interactions might be buried in this symmetry. 
In the following I want to extend this analysis in several 
directions. First, I will show that it is possible to 
embed the discrete $S_4$-symmetry in a larger 
continuous symmetry group. Afterwards, we shall see how 
the appearance of gauge bosons can be understood as well as 
obtain some hints about how the underlying microscopic structure 
may look like. 

The permutation group $S_4$ \cite{s4} consists of 5 classes with altogether 24 
elements $\sigma = \overline{abcd}$ where $a,b,c,d \in \{1,2,3,4\}$. 
It has 5 representations $A_1$, $A_2$, $E$, $T_1$ and $T_2$ of 
dimensions 1, 1, 2, 3 and 3 and is isomorphic to the 
symmetry group $T_d$ of a regular tetrahedron (and also 
to the subgroup $O$ of proper rotations of the symmetry group 
$O_h$ of a cube), cf table 1. 
The observed fermion symmetry will therefore be synonymously 
called $T_d$ or $S_4$ in the following, depending on whether 
a geometrical or an algebraic viewpoint is taken. 

An important subgroup of $S_4$ is $A_4$, the group of 
even permutations, which is sometimes called the 'symmetric group' 
and will be relevant in the discussion of gauge bosons in 
section 5. $A_4$ has 3 representations $A$, $E$ and $T$ of 
dimensions 1, 2 and 3 and is isomorphic to the 
symmetry group of proper rotations of a regular 
tetrahedron. 

\begin{table}
\label{tab1}
\begin{center}
\begin{tabular}{|l|c|c|c|}
\hline
& $S_4$ & $T_d$ & $O$ \\
\hline
I & $\overline{1234} (id)$ & identity-rotation & identity-rotation \\
\hline
3$C_2$ & $\overline{2143}$ $\overline{3412}$ $\overline{4321}$ & rotations by $\pi$ about & rotation by $\pi$ about \\
       &                                         & the coordinate axes & the coordinate axes \\
\hline
8$C_3$ & $\overline{2314}$ $\overline{3124}$ $\overline{3241}$ $\overline{1342}$ & rotations by $\frac{2}{3}\pi$ about & rotations by $\frac{2}{3}\pi$ about  \\
       & $\overline{1423}$ $\overline{2431}$ $\overline{4132}$ $\overline{4213}$ & diagonals of the cube           & diagonals of the cube \\
\hline
6$C_4$ & 6 transpositions                            & 6 reflections on planes         & rotations by $\pm\frac{1}{2}\pi$ about\\
       & $(i\leftrightarrow j)$ like                 & through the center              & the coordinate axes \\
       & $(1\leftrightarrow 2)=\overline{2134}$      & and two edges i and j           &  \\
\hline
6$C_2'$ & $\overline{2341}$ $\overline{3142}$  & 6 rotoreflections by $\frac{1}{2}\pi$  & rotations by $\pi$ about \\
        & $\overline{2413}$ $\overline{3421}$  &                                        & axes parallel to the 6 \\ 
        & $\overline{4123}$ $\overline{4312}$  &                                        & face diagonals \\
\hline
\end{tabular}
\bigskip
\caption{Classes I, $C_2$, $C_3$, $C_4$ and $C_2'$ of the 
groups $S_4$, $T_d$ and $O$ making their isomorphy explicit. 
Classes I, $C_2$ and $C_3$ form the 12-element subgroup $A_4$ of 
even permutations, which will be important in our analysis 
of vector bosons in section 5. The notation $C_4$ and $C_2'$ 
is normally used only for rotations in $O$, whereas the 
classes of reflections in $T_d$ are usually called $\sigma$ 
and $S_4$ in the literature.}
\end{center}
\end{table}

The starting point of ref. \cite{lampe2} was the observation that 
there is a natural one-to-one correspondence between the 
fermion states and the elements of $S_4$. This feature is made 
explicit in table 2 where the elements of $S_4$ 
are associated to the fermions. 

\begin{table}
\label{tab2}
\begin{center}
\begin{tabular}{|l|c|c|c|}
\hline
&...1234...&...1423...&...1243... \\
& family 1 & family 2 & family 3 \\
\hline
& $\tau$, $b_{1,2,3}$ & $\mu$, $s_{1,2,3}$ & $e$, $d_{1,2,3}$ \\
\hline
$\nu$ & $\overline{1234} (id)$ & $\overline{2314}$ & $\overline{3124}$ \\
$u_1$ & $\overline{2143} (k_1)$ & $\overline{3241}$ & $\overline{1342}$ \\
$u_2$ & $\overline{3412} (k_2)$ & $\overline{1423}$ & $\overline{2431}$ \\
$u_3$ & $\overline{4321} (k_3)$ & $\overline{4132}$ & $\overline{4213}$ \\
\hline
& $\nu_\tau$, $t_{1,2,3}$ & $\nu_{\mu}$, $c_{1,2,3}$  & $\nu_e$, $u_{1,2,3}$  \\
\hline
$l$ & $\overline{3214} (1\leftrightarrow 3)$ & $\overline{1324} (2\leftrightarrow 3)$ & $\overline{2134} (1\leftrightarrow 2)$ \\
$d_1$ & $\overline{2341}$ & $\overline{3142}$ & $\overline{1243} (3\leftrightarrow 4)$ \\
$d_2$ & $\overline{1432} (2\leftrightarrow 4)$ & $\overline{2413}$ & $\overline{3421}$ \\
$d_3$ & $\overline{4123} $ & $\overline{4231} (1\leftrightarrow 4)$ & $\overline{4312}$ \\
\hline
\end{tabular}
\bigskip
\caption{List of elements of $S_4$ ordered in 3 families. 
$k_i$ denote the elements of K and $(a\leftrightarrow b)$ a 
simple permutation where a and b are interchanged. 
Permutations with a 4 at the last position form 
a $S_3$ subgroup of $S_4$ and may be thought of giving the set 
of lepton states. 
It should be noted that this is only a heuristic 
assignment. Actually one has to consider linear combinations 
of permutation states as discussed in section 2.}
\end{center}
\end{table}

I use the term 'natural' because the color, isospin and 
family structure of fermions corresponds to K, $Z_2$ 
and $Z_3$ subgroups of $S_4$, where $Z_n$ is the (abelian) 
symmetric group of n elements and K is the 
so-called Kleinsche Vierergruppe which 
consists of the 3 even permutations $\overline{2143}$, 
$\overline{3412}$, $\overline{4321}$, where 2 pairs of 
numbers are interchanged (class $C_2$), plus the identity. 
In fact, $S_4$ is a semi-direct product 
$S_4= K \diamond Z_3 \diamond Z_2$ where the 
$Z_3$ factor is the family symmetry and $Z_2$ and K can 
be considered to be the 'germs' of weak isospin and color 
(cf \cite{lampe2} and section 5). 
At low energies this product cannot be distinguished 
from the direct product $K \times Z_3 \times Z_2$ 
but has the advantage of being a simple group and having a 
rich geometric and group theoretical interpretation and 
will also lead to a new ordering scheme for the Standard 
Model vector bosons in section 5. 

If one wants to include antiparticles and the spin of the fermions 
in this analysis, one can do the following: relativistically the 
situation seems very simple. Spin and antiparticles 
each double the degrees of freedom, so that one has the structure of table 2 
for $f_L$, $f_R$, $\bar f_L$ and $\bar f_R$ separately. 
This is enough, as long as one continues to consider quarks and 
leptons as pointlike objects, and asks questions like how under 
the assumption of the $S_4$ symmetry vector boson formation 
can be interpreted (section 5), and as long as one keeps the 
(discrete) inner and spatial symmetries completely separate -  
but it would not suffice any more, as soon as one would 
consider the possibility of compositeness and a spatial 
extension of the observed fermions, in particular in the 
form of a micro-geometric tetrahedral substructure 
\cite{lampef},\cite{lampe2}. 

In that case the situation becomes much more difficult. 
The point is that a tetrahedron is not relativistically  
invariant and one does not have a relativistic 
description of such an extended object. As an alternative 
one may try \cite{lampe2} to use a nonrelativistic approach 
to spin and antiparticles by going from $T_d$ to $\tilde O_h$, 
which is the covering group of the octahedral group $O_h$. 
$O_h$ is in fact just the direct product $T_d\times P$, 
where P is the space inversion symmetry. 
Going from $T_d$ to $\tilde O_h$ amounts more or 
less to adding 2 factors of $Z_2$ to $T_d$, one corresponding to 
spin and one for antiparticles (complex conjugation). 
In addition to the ordinary representations one then 
has to include the representation $G_1$ of the covering 
group \cite{johnson}.
As can be shown, this amounts to introducing 
two functions $f^+_{\sigma}$ and $f^-_{\sigma}$ 
where the spin averaged wave function is given by the sum 
\begin{equation} 
f_{\sigma}=f^+_{\sigma}+f^-_{\sigma}
\label{eq1sd}
\end{equation}
whereas the spin content is contained in the difference 
$f^+_{\sigma} - f^-_{\sigma}$, 
and means that including the spin degrees of freedom one has 
now 48 wave functions instead of the 24 given in table 1. 

One may visualize this approach by a geometrical picture, 
where one has a cube which contains two tetrahedra 
(one for particles and the other one for antiparticles) 
which transform into each other by a CP-transformation so 
that for example in the process of vector boson formation 
$\bar F \gamma_\mu f$ the fermion f, which spreads over 
the first tetrahedron, and antifermion $\bar F$, which 
spreads over the other, join together to form a cube. 

It should be noted that even if one rejects the 
constituent and spatial extension picture it is 
possible to give a meaning to the tetrahedra 
describing $f_L$ and $f_R$ and being connected 
by parity. For example, in the SU(4) model 
which will be introduced in section 4, 
they do not live in physical space but 
exist as weight diagrams of the fundamental SU(4) 
representation. If one follows such an approach 
(which will be done for the most part of the paper) 
a correct relativistic treatment can be maintained without 
any difficulty. 


\section{The Use of Symmetry adapted Wave Functions and 
         the Origin of strong and electroweak Charges} 


In \cite{lampe2} a sort of seesaw mechanism was derived 
which is able to accomodate all observed  
hierarchies in the quark and lepton masses. 
This mechanism relies on the introduction of 
$S_4$ symmetry functions to describe fermion fields, where 
the given Dirac fields of quarks and leptons are written as 
{\bf symmetry adapted linear combinations} of more fundamental 
fields $\psi_{\sigma}$, $\sigma \in S_4$. 

The linear coefficients are essentially given by the 
$A_1$, $A_2$, $E$, $T_1$ and $T_2$ representation matrices 
of $S_4$. This is due to the group 
theoretic theorem that from an arbitrary function $f(x)$ orthonormal 
sets of symmetry functions of a discrete group G can be obtained as 
\begin{eqnarray}  
f_{ij}=\frac{dim(D)}{|G|}\sum_{g\in G} D_{ij}(g)f(g^{-1}x)
\label{hhg145}
\end{eqnarray} 
where D is any representation of G. 
(In general this will yield dim(D) sets of dim(D) orthonormal 
symmetry functions corresponding to the representation D.)
Therefore to obtain the symmetry adapted functions one just 
has to take as linear coefficients 
the appropriate representation matrix entries $D_{ij}$ 
which are well known in the realm of finite symmetry 
groups and for convenience given in tables 3 and 4 \cite{taylor1}. 
The resulting functions were already given in ref. \cite{lampe2}.

\begin{table}
\label{tab331}
\begin{center}
\begin{tabular}{|l|c|c|c|c|c|c|c|c|c|c|c|c|}
\hline
   & &x    &  y&z    &xyz  &$\bar x y \bar z$&$\bar x \bar y z$&$x\bar y \bar z$& xyz & $\bar x y \bar z$&$\bar x\bar y z$& $x\bar y\bar z$\\
   &1&$C_2$&$C_2$&$C_2$&$C_8$            &$C_8$            &$C_8$           &$C_8$&$C_8$             &$C_8$           &$C_8$          &$C_8$\\ 
\hline
$A_1$       & 1&  1&  1&  1&  1&  1&  1&  1&  1&  1&  1&  1  \\
$A_2$       & 1&  1&  1&  1&  1&  1&  1&  1&  1&  1&  1&  1  \\
$(E)_{11}$  & 1&  1&  1&  1&  c&  c&  c&  c&  c&  c&  c&  c  \\
$(E)_{21}$  & 0&  0&  0&  0&  s&  s&  s&  s &-s &-s &-s &-s  \\ 
$(E)_{12}$  & 0&  0&  0&  0 &-s &-s &-s &-s&  s&  s&  s&  s  \\ 
$(E)_{22}$  & 1&  1&  1&  1&  c&  c&  c&  c&  c&  c&  c&  c  \\
$(T_1)_{11}$& 1&  1 &-1 &-1&  0&  0&  0&  0&  0&  0&  0&  0  \\ 
$(T_1)_{21}$& 0&  0&  0&  0&  1 &-1&  1 &-1&  0&  0&  0&  0  \\
$(T_1)_{31}$& 0&  0&  0&  0&  0&  0&  0&  0&  1 &-1 &-1&  1  \\
$(T_1)_{12}$& 0&  0&  0&  0&  0&  0&  0&  0&  1&  1 &-1 &-1  \\
$(T_1)_{22}$& 1 &-1&  1 &-1&  0&  0&  0&  0&  0&  0&  0&  0  \\
$(T_1)_{32}$& 0&  0&  0&  0&  1 &-1 &-1&  1&  0&  0&  0&  0  \\
$(T_1)_{13}$& 0&  0&  0&  0&  1&  1 &-1 &-1&  0&  0&  0&  0  \\
$(T_1)_{23}$& 0&  0&  0&  0&  0&  0&  0&  0&  1 &-1&  1 &-1  \\
$(T_1)_{33}$& 1 &-1 &-1&  1&  0&  0&  0&  0&  0&  0&  0&  0  \\
$(T_2)_{11}$& 1&  1 &-1 &-1&  0&  0&  0&  0&  0&  0&  0&  0  \\
$(T_2)_{21}$& 0&  0&  0&  0&  1 &-1&  1 &-1&  0&  0&  0&  0  \\
$(T_2)_{31}$& 0&  0&  0&  0&  0&  0&  0&  0&  1 &-1 &-1&  1  \\
$(T_2)_{12}$& 0&  0&  0&  0&  0&  0&  0&  0&  1&  1 &-1 &-1  \\
$(T_2)_{22}$& 1 &-1&  1 &-1&  0&  0&  0&  0&  0&  0&  0&  0  \\
$(T_2)_{32}$& 0&  0&  0&  0&  1 &-1 &-1&  1&  0&  0&  0&  0  \\ 
$(T_2)_{13}$& 0&  0&  0&  0&  1&  1 &-1 &-1&  0&  0&  0&  0  \\
$(T_2)_{23}$& 0&  0&  0&  0&  0&  0&  0&  0&  1 &-1&  1 &-1  \\
$(T_2)_{33}$& 1 &-1 &-1&  1&  0&  0&  0&  0&  0&  0&  0&  0  \\
\hline
\end{tabular}
\bigskip
\caption{Matrices for the irreducible representations of $S_4=T_d$ fixing the coefficients 
of the symmetry adapted functions as given in \cite{taylor1}. I have used the abbreviation 
$c = cos(\frac{2}{3}\pi) =-\frac{1}{2}$ and 
$s = sin(\frac{2}{3}\pi) =\frac{\sqrt{3}}{2}$. 
}
\end{center}
\end{table}

\begin{table}
\label{tab332}
\begin{center}
\begin{tabular}{|l|c|c|c|c|c|c|c|c|c|c|c|c|}
\hline
   &  $\bar x y$& xy& $\bar x z$& xz &$\bar y z$& yz& z& z& y& y& x& x \\
   &$\sigma$&$\sigma$&$\sigma$&$\sigma$&$\sigma$&$\sigma$&$S_4$&$S_4$&$S_4$&$S_4$&$S_4$&$S_4$\\ 
\hline
$A_1$       &  1&  1&  1&  1&  1&  1&  1&  1&  1&  1&  1&  1 \\
$A_2$       & -1 &-1 &-1 &-1 &-1 &-1 &-1 &-1 &-1 &-1 &-1 &-1 \\
$(E)_{11}$  & 1&  1&  c&  c&  c&  c&  1&  1&  c&  c&  c&  c  \\
$(E)_{21}$  & 0&  0&  s&  s &-s &-s&  0&  0&  s&  s &-s &-s \\ 
$(E)_{12}$  & 0&  0&  s&  s &-s &-s&  0&  0&  s&  s &-s &-s \\ 
$(E)_{22}$  & -1 &-1 &-c &-c &-c &-c &-1 &-1 &-c &-c &-c &-c \\
$(T_1)_{11}$& 0&  0&  0&  0 &-1 &-1&  0&  0&  0&  0&  1&  1 \\ 
$(T_1)_{21}$& -1&  1&  0&  0&  0&  0 &-1&  1&  0&  0&  0&  0 \\
$(T_1)_{31}$& 0&  0 &-1&  1&  0&  0&  0&  0&  1 &-1&  0&  0 \\
$(T_1)_{12}$& -1&  1&  0&  0&  0&  0&  1 &-1&  0&  0&  0&  0 \\
$(T_1)_{22}$& 0&  0 &-1 &-1&  0&  0&  0&  0&  1&  1&  0&  0 \\
$(T_1)_{32}$& 0&  0&  0&  0 &-1&  1&  0&  0&  0&  0 &-1&  1 \\
$(T_1)_{13}$& 0&  0 &-1&  1&  0&  0&  0&  0 &-1&  1&  0&  0 \\
$(T_1)_{23}$& 0&  0&  0&  0 &-1&  1&  0&  0&  0&  0&  1 &-1 \\
$(T_1)_{33}$& -1 &-1&  0&  0&  0&  0&  1&  1&  0&  0&  0&  0 \\
$(T_2)_{11}$& 0&  0&  0&  0&  1&  1&  0&  0&  0&  0 &-1 &-1 \\
$(T_2)_{21}$& 1 &-1&  0&  0&  0&  0&  1 &-1&  0&  0&  0&  0 \\
$(T_2)_{31}$& 0&  0&  1 &-1&  0&  0&  0&  0 &-1&  1&  0&  0 \\
$(T_2)_{12}$& 1 &-1&  0&  0&  0&  0 &-1&  1&  0&  0&  0&  0 \\
$(T_2)_{22}$& 0&  0&  1&  1&  0&  0&  0&  0 &-1 &-1&  0&  0 \\
$(T_2)_{32}$& 0&  0&  0&  0&  1 &-1&  0&  0&  0&  0&  1 &-1 \\ 
$(T_2)_{13}$& 0&  0&  1 &-1&  0&  0&  0&  0&  1 &-1&  0&  0 \\
$(T_2)_{23}$& 0&  0&  0&  0&  1 &-1&  0&  0&  0&  0 &-1&  1 \\
$(T_2)_{33}$& 1&  1&  0&  0&  0&  0 &-1 &-1&  0&  0&  0&  0 \\
\hline
\end{tabular}
\bigskip
\caption{Continuation of table 3: representation matrices for the reflection 
operations in $T_d$. The symbols above the symmetry operations indicate their 
orientation relative to the axes.}
\end{center}
\end{table}

In order to explain the observed 
parity violation of the weak {\it and} the $V-A$ structure of the 
strong interaction it was suggested \cite{lampe2} 
that the two tetrahedra describing fermions and antifermions 
are intertwined in the following sense: field components 
$\psi_g$ corresponding to even permutations $g \in S_4$ 
live on one tetrahedron, whereas components $\psi_u$ corresponding 
to odd permutations $u \in S_4$ live on the other. In other words, 
the symmetry adapted functions for left handed fermions have the 
generic form $f_L=\psi_g+P\psi_u$ and those for the right handed  
$f_R=P \psi_g+ \psi_u$. The point is that 
fermions of opposite isospin differ by an odd permutation 
(as is explicit from table 2), so that parity violation/conservation 
for weak bosons/gluons is obtained.\cite{lampe2} 

Having made extensive use of symmetry adapted functions 
in various directions, it is time to discuss the legitimacy and 
drawbacks of such an approach, which have to do with the fact 
that one is combining fields with different Standard Model 
charges into linear combinations. 
As a consequence no definite strong and electroweak charges can 
be associated to single state componenents $\psi_{\sigma}$, 
$\sigma=\overline{abcd} \in S_4$, but only 
to the symmetry adapted linear combinations giving the 
quarks and leptons.  
In other words, such an approach can only be valid, if 
{\bf the Standard Model charges arise as derived entities from 
secondary dynamical causes and are not really fundamental.} 
Fundamental are only the interactions behind the 
$S_4$-symmetry (resp. SU(4)-symmetry in section 4) or 
the superstrong forces between the possible constituents, 
whereas the Standard Model interactions of the fermions 
do not exist a priori but are just a consequence of the 
relative position of a, b, c and d in the permutations. 
In order to understand this more clearly it was 
suggested in \cite{lampe2} to introduce nondiagonal 
charge operators so that not the permutation fields $\psi_{\sigma}$ 
but their symmetry combinations are eigenfunctions of the Standard Model 
charge operators - in much the same way as they are not eigenfunctions 
of the mass operator. 


If one does not like this approach and 
wants to stick to the viewpoint that charge operators must be  
diagonal and have to be associated not to linear combinations of 
fields but to the fields $\psi_{\sigma}$ themselves, one has to 
give up the symmetry adapted linear combinations. 
The only linear combinations which may then be used are $Z_3$-adapted 
functions, because they are not associated to any charges but to the 
family symmetry. In other words, since for example the 3 
neutrinos, for which permutations of the first 3 indices are 
relevant (cf table 2), have identical Standard Model 
charges, one may use linear combinations of the form 
\begin{eqnarray}  
\nu_e  &=&\psi_{\overline{1234}}+\psi_{\overline{2314}}+\psi_{\overline{3124}} \label{eqzz3} \\
\nu_{\mu}&=&\psi_{\overline{1234}}+\epsilon\psi_{\overline{2314}}+\epsilon^*\psi_{\overline{3124}} \\
\nu_\tau&=&\psi_{\overline{1234}}+\epsilon^*\psi_{\overline{2314}}+\epsilon\psi_{\overline{3124}} 
\end{eqnarray} 
and similarly for electron, muon and tau-lepton 
\begin{eqnarray}  
e  & =&\psi_{\overline{3214}}+\psi_{\overline{1324}}+\psi_{\overline{2134}} \\
\mu& =&\psi_{\overline{3214}}+\epsilon\psi_{\overline{1324}}+\epsilon^*\psi_{\overline{2134}} \\
\tau&=&\psi_{\overline{3214}}+\epsilon^*\psi_{\overline{1324}}+\epsilon\psi_{\overline{2134}} 
\label{eqzz4}
\end{eqnarray} 
These equations are easily understood because $Z_3$-symmetry 
combinations always have the generic 
form $f_0+f_1+f_2$, $f_0+\epsilon f_1+\epsilon^*f_2$ 
and $f_0+\epsilon^* f_1+\epsilon f_2$, 
where $\epsilon=exp(2\pi i/3)$. 

Gauge bosons may be re-expressed using these combinations. 
For example one obtains for the leptonic part of the neutral weak W-boson 
\begin{eqnarray}  
W_{3\mu}&=&\bar e \gamma_\mu e - \bar\nu_e \gamma_\mu \nu_e            
     +\bar \mu \gamma_\mu \mu - \bar\nu_\mu \gamma_\mu \nu_\mu      
     +\bar \tau \gamma_\mu \tau - \bar\nu_\tau \gamma_\mu \nu_\tau              \\
     &=&3( \bar \psi_{\overline{1234}} \gamma_\mu \psi_{\overline{1234}} 
          +\bar \psi_{\overline{2314}} \gamma_\mu \psi_{\overline{2314}}
          +\bar \psi_{\overline{3124}} \gamma_\mu \psi_{\overline{3124}}\nonumber \\ 
     & &  -\bar \psi_{\overline{3214}} \gamma_\mu \psi_{\overline{3214}}
          -\bar \psi_{\overline{1324}} \gamma_\mu \psi_{\overline{1324}}
          -\bar \psi_{\overline{2134}} \gamma_\mu \psi_{\overline{2134}})
\label{eqzzx5}
\end{eqnarray} 

Note that eqs. (\ref{eqzz3})-(\ref{eqzzx5}) hold separately for left and right handed 
lepton and W fields.


\section{The two main Problems} 

In the remainder of this work I will deal with the two 
fundamental problems, which have to be solved, if the tetron 
approach is to make sense: 
\begin{itemize}
\item 
First to understand in a natural way the appearance of vector 
bosons 
as linear combinations of products of fermion fields. 
In particular the question why among the many fermion-antifermion products 
which can in principle be formed, precisely and only those corresponding 
to the Standard Model gauge groups arise. 
The idea which reduces the number of possible combinations and 
produces the Standard Model gauge 
bosons will be that when product states are formed from 
two fermions each with $T_d$- resp. $O_h$-symmetry 
a final state object appears, which again has a symmetry 
of (a subgroup of) $T_d$.  
\item 
Secondly what the underlying origin of the tetrahedral symmetry may be. 
It is plausible although not compelling that the observed $S_4$-symmetry 
points to a substructure of quarks and leptons with four constituents. 
In this scenario the main question is how the spin-$\frac{1}{2}$ 
nature of the observed fermions can be obtained. 
One possibility, which will be followed in a separate 
publication\cite{lampef}, is to give up continuous spatial 
rotation symmetry on the microscopic level and replace it 
by a discrete (tetrahedral or octahedral) symmetry and then 
to consider $Z_4$-extensions of the tetrahedral group instead 
of the $Z_2$-extension defined by the covering group. 
There is then the possibility that for this $Z_4$-extension 
quaternion instead of complex quantum mechanics may 
play a role. 
\end{itemize}

I consider the first problem more important, in particular in view 
of the highly speculative nature of the second one. 


\section{Discrete versus continuous inner Symmetry} 

I have repeatedly mentioned the argument of ref.\cite{lampe2} 
that $S_4$-symmetry 
transformations may serve as 'germs' for the gauge 
symmetries which in modern times are used to 
describe the strong and electroweak interactions. 

Discrete symmetry as an ordering scheme for quarks 
and leptons and a possible source for their interactions? 
At this point particle physicists may 
feel a bit uneasy, because it can hardly be imagined 
that the rich and rather involved structure of the Standard 
Model gauge theories can be derived in a strict sense 
from a discrete symmetry structure.  

Therefore, one may look for alternative ideas, and one possibility 
is that the appearant $S_4$-symmetry of quarks and leptons is 
part of a larger (continuous) symmetry group like SU(4) or Sp(4). 
In these groups the $S_4$-symmetry adapted functions naturally 
appear as part of the product states in 
$\bf 4\otimes \bf 4\otimes \bf 4\otimes \bf 4$,  
where {\bf 4} is the fundamental representation of SU(4), 
the representation space being spanned by 'tetron' states 
a, b, c and d, just like in the $SU(3)_{flavor}$ quark model 
the fundamental representation {\bf 3} is spanned by fields u, d and s. 
The point is that if one considers fourfold tensor products 
$\bf 4\otimes \bf 4\otimes \bf 4\otimes \bf 4$, 
among the corresponding 256 possible states one will automatically 
encounter the 24 linear combinations of product states 
$\psi_{\overline{abcd}}=a\times b\times c\times d$ and their permutations, 
or more precisely the symmetry adapted linear combinations 
thereof - just like in the $SU(3)_{flavor}$ quark model 
among the 27 baryonic states in $\bf 3\otimes \bf 3\otimes \bf 3$ 
there are 6 linear combinations like for example 
$\Lambda^0=\frac{1}{\sqrt{12}}[sdu-sud+usd-dsu+2(uds-dus)]$ which 
can be interpreted as symmetry adapted functions of the 
permutation group $S_3$. 
This is not astonishing but has to do with the fact 
that $S_4$($S_3$) is a distinct particle symmetry of the product 
states in $\bf 4\otimes \bf 4\otimes \bf 4\otimes \bf 4$
($\bf 3\otimes \bf 3\otimes \bf 3$). 

Since the fundamental representation of SU(4) can be 
geometrically visualized as a tetrahedron which 
lives in a 3-dimensional weight diagram spanned by the 
SU(4) generators $\lambda_{3,8,15}$, 
we arrive at more or less the same geometrical picture as 
described in section 1 for the discrete $S_4$-symmetry. 
Even the formation of vector bosons as compounds 
$\bar F \gamma_\mu f$ from two tetrahedral configurations, 
which can be transformed into another by CP 
and where a cube is formed in the combined 
weight diagram of particles and antiparticles, 
can be understood in this model. 

There are 3 questions left open: 
\begin{itemize}
\item 
how the Standard Model charges and interactions can arise 
from an SU(4) 'hyperflavor' interaction just by a permutation 
of constituents. This question will be tackled in section 5.  
\item 
how products of 4 constituents can make up for 
fermions with their spin-$\frac{1}{2}$ transformation properties 
under spatial rotations. This will be discussed in 
a forthcoming publication \cite{lampef}. 
\item 
and finally why only 'distinct'-tetron 
states arise, whereas all the rest of the 256 product states 
(those where one of the tetrons appears at least twice) 
are not observed (or have a much higher mass). 
\end{itemize}

As for the last problem I formulate the following 
{\bf exclusion principle for tetrons}: quarks and leptons 
consist of 4 tetron states a,b,c,d. 
Only states where all tetrons are different are allowed. 
In order to include vector bosons and their treatment 
in section 5 one may extend this principle as follwows:  
for an arbitrary state to be physical 
the exclusion principle demands that it is part 
of a $S_4$ permutation multiplet. 

Note that this is a weaker condition (i.e. allows more 
states) than for example the color singlet principle 
of $SU(3)_{color}$-QCD, which demands that among all 
$\bf 3\otimes \bf 3\otimes \bf 3$ only the $A_2$ 
singlet function $\epsilon(i,j,k)q_i q_j q_k$ is allowed. 

In conclusion one may say that one has two options which match 
the phenomenological fermion spectrum equally well: 
either one uses a continuous inner symmetry group like SU(4) 
$together$ $with$ an exclusion principle or one sticks 
to the discrete tetrahedral=permutation symmetry.  

One can make the connection between these two approaches 
explicit by writing down the $T_d$-content of the relevant SU(4) 
representations. Namely, within the discrete approach the 
24 fermion states can be classified according to the 
$T_d$ representations $A_1$, $A_2$, $E$, $T_1$ and $T_2$, 
i.e. the 18 $T_1$- and $T_2$-functions are used to describe up- and 
down-type quarks degrees of freedom respectively, whereas the 
6 $A_1$-, $A_2$- and $E$-functions are responsible for leptons. 
(This is just the use of the symmetry adapted functions 
discussed before and in \cite{lampe2}.) 
On the other hand, in the continuous symmetry approach 
the 256 SU(4)-states of $\bf 4\otimes \bf 4\otimes \bf 4\otimes \bf 4$ 
may be decomposed according to 
\begin{equation} 
{\bf 4\otimes 4\otimes 4\otimes 4}=3{\bf \times 45(T_1)}+3{\bf \times 15(T_2)}
                 +2{\bf \times 20(E)}+{\bf 35(A_1)}+{\bf 1(A_2)}
\label{eq1sdd}
\end{equation}
Here one finds in brackets, which kind of $T_d$ symmetry functions are 
contained in the corresponding SU(4) representations. 
For example, there are three SU(4) representations of dimension 45 
each containing a set of 3 $T_1$-functions, i.e. all in all the 
9 functions used to describe the up-type quarks. 
More precisely, the 3 functions of the n-th $T_1$ in 
(\ref{eq1sdd}) are to describe the family triplet 
$u_n$, $c_n$ and $t_n$, where n=1,2,3 is the color index. 
Similarly there are 3 sets of 3 $T_2$-functions in the 3 
15-dimensional representations to describe the down-type quarks. 
Furthermore, $A_1$ and $A_2$ describes the electron and its neutrino, 
whereas one E-representation in (\ref{eq1sdd}) 
contains $\mu$ and $\tau$ and the other 
$\nu_\mu$ and $\nu_\tau$. 

It is an interesting observation that 
this way only particles of the same Standard Model charges 
(but belonging to different families) are put together in a SU(4) multiplet. 
The alternative would be to put quarks of different color into 
one SU(4) multiplet (like $u_1$, $u_2$ and $u_3$ into one ${\bf 45}$) 
and similarly for leptons of different isospin (e.g. $\mu$ and $\nu_\mu$ 
into one ${\bf 20}$). 

It should further be noted that the fermion mass relations derived 
in \cite{lampe2} on the basis of the discrete $T_d$-symmetry can be 
rederived as SU(4) mass relations that are analogous to the mass 
relations for hadrons derived in the $SU(3)_{flavor}$ quark model. 


\section{Vector Boson Formation} 

In this section I will not make any assumptions about 
possible substructures of quarks and leptons, but will 
only use the appearant $S_4$-symmetry of their spectrum table 2. 
On the basis of this symmetry I want to show that the vector 
bosons of the left-right symmetric Standard Model 
can be ordered in a similar manner and according to the 
same principle as the fermions. 
The idea is that the tetrahedral (resp octahedral) symmetry 
of the quarks and leptons is more or less retained when the vector 
bosons are formed. More precisely, I shall assume 
that the vector boson states can be ordered according 
to the subgroup $A_4$ of $S_4$ (the so called symmetric 
group of even permutations). 
This reduces the a priori large 
number of possible fermion-antifermion interactions, 
because it means that whatever internal dynamical reordering takes place 
in the process of vector boson formation $\bar F \gamma_\mu f$ 
from two fermions F and f, the resulting state has to have 
$A_4$ symmetry. 
For example, the long discussion in ref.\cite{lampe2} of 
how to avoid leptoquarks is completely superfluous in this 
approach simply because within the $A_4$-symmetry with its 
12 degrees of freedom there is no space for additional 
gauge bosons.

The two possible types of vector bosons 
$V_{\mu L}=\bar F_L \gamma_\mu f_L$ and 
$V_{\mu R}=\bar F_R \gamma_\mu f_R$ 
can be accounted for by including 
parity $P:V_{\mu L} \leftrightarrow V_{\mu R}$ 
so that one arrives at the 
so called pyritohedral symmetry $A_4\times P$, a 
subgroup of the octahedral group $O_h$. 
Note that since the gauge bosons have spin 1, no 
covering group has to be considered. 
Note further that since I work in the relativistic limit 
(which I can do since $S_4$ and $A_4$ 
are just inner symmetries of pointlike particles)
no vector boson spin-0 component appears.  
 
In table 5 I present a heuristic ordering of the observed vector 
bosons according to the proposed $A_4$-symmetry. Phenomenologically, 
there are 8 gluons $G_\mu$, one $(B-L)$-photon $B_\mu$ and 3 
weak bosons $W_{1,2,3\mu}$. 
The argument of why only the weak 
bosons appear in a right- and a lefthanded version $W_R$ 
and $W_L$, whereas for gluons and photon one has 
$G_{\mu L}=G_{\mu R}$ and $B_{\mu L}=B_{\mu R}$ 
can be taken over from ref \cite{lampe2}.

\begin{table}
\label{tab9}
\begin{center}
\begin{tabular}{|l|c|c|}
\hline
$B_{\mu}=\overline{1234}$    & $G_{3\mu}=\overline{2314}$ & $G_{8\mu}=\overline{3124}$ \\
$W_{3\mu}=\overline{2143}$   & $G_{1\mu}=\overline{3241}$ & $G_{2\mu}=\overline{1342}$ \\
$W_{1\mu}=\overline{3412}$   & $G_{4\mu}=\overline{1423}$ & $G_{5\mu}=\overline{2431}$ \\
$W_{2\mu}=\overline{4321}$   & $G_{6\mu}=\overline{4132}$ & $G_{7\mu}=\overline{4213}$ \\
\hline
\end{tabular}
\bigskip
\caption{List of Standard Model vector bosons 
ordered heuristically according to their proposed $A_4$ symmetry. 
$A_4$ is composed of 3 classes I, 3$C_2$, 8$C_3$ (cf table 1) 
and the proposed ordering follows this line. Note that just as 
table 2 for fermions these 
are only preliminary assignments. Later we shall see, how 
to construct the correct vector bosons states in terms 
of symmetry adapted functions.}
\end{center}
\end{table}

This table, which may look miraculous at first sight, is not 
difficult to understand. For example, in \cite{lampe2} it was 
argued that the weak bosons 
$W_{1,2,3}$ arise naturally from the Kleinsche Vierergruppe 
K (the subgroup of $A_4$ formed by the classes I and 
3$C_2$) because it is isomorphic to $Z_2\times Z_2$ where 
the two $Z_2$ factors stand for the germs of weak isospin 
of the fermion resp antifermion. 

To go beyond such a heuristic understanding one should use symmetry 
adapted linear combinations of functions $\Psi_\sigma$, $\sigma \in A_4$ 
instead of the simple assignments of table 5. 
The linear coefficients could in principle be taken from table 3 
(dropping the contributions from improper rotations). 
However we shall instantly see how to construct them 
explicitly from fermion-antifermion bilinears in order to 
obtain the combinations relevant in particle physics.

Using $S_4$-Clebsch-Gordon coefficients 
for the fermion-antifermion tensor products \cite{griff}, 
I want to show, that and how from the 24x24=512 
possible fermion-antifermion-product states 12 are selected 
in order to describe the final states (the vector bosons). 
From the point of principle this is in fact no question: 
if the final states are to have $A_4$-symmetry 
then their number {\it must} boil down to 12. 
In practice these states can be explicitly 
constructed by evaluating fermion-antifermion products using 
the $S_4$-symmetry adapted functions for the fermions whose 
benefits and deficiencies have been discussed in section 2, 
also in connection with their appearance in the 
continuous SU(4) model in section 4, cf. eq. (\ref{eq1sdd}), 
projecting them to $A_4\subset S_4$ and comparing the result 
with the observed vector boson spectrum. 

Before I start I want to remind the reader that the 24 $S_4$-functions 
for fermions divide into 9 symmetry functions from $T_1$ used for the 
up-type-quarks, 9 functions from $T_2$ for the down-type-quarks 
and 6 functions from $A_1$, $A_2$ and E for the lepton degrees 
of freedom and that they all can be obtained from table 3. 
Clebsch-Gordon(CG) coefficients appear when one calculates tensor 
products of two representations $D_1$ and $D_2$ 
as direct sums 
\begin{equation}
D_1\otimes D_2=D_3\oplus ... 
\label{eqt11253}
\end{equation}
and wants to determine a set of symmetry functions for $D_3$ from 
symmetry functions $f_1^i$ and $f_2^j$ of $D_1$ and $D_2$. 
Namely they are given 
\begin{equation} 
f_3^k=\sqrt{dim(D_3)}\sum_{i,j}V(D_1,D_2,D_3,i,j,k)f_1^if_2^j
\label{eqt153}
\end{equation}
where the sum runs over sets of symmetry functions that span 
the representation spaces, $i=1,...,dim(D_1)$ and $j=1,...,dim(D_2)$.
Eq. (\ref{eqt153}) will be used as the defining equation 
for the normalization of the CG-coefficients. 
(In fact we are using so-called V-coefficients which 
have the advantage of being invariant under 
simultaneous permutations of representations 
and indices in their argument.)

Consider for example the product $T_1\otimes T_1$. Since $T_1$ 
corresponds to the up-type quarks, the product $T_1\otimes T_1$ will 
yield 9 up-quark bilinears. Within $S_4$ these can be decomposed according to 
\begin{equation} 
T_1\otimes T_1 = A_1 \oplus E \oplus T_1 \oplus T_2
\label{eqt1}
\end{equation}
Taking the 3 up-quark color components $u_1$, $u_2$ and $u_3$ 
as $T_1$-functions on the LHS and evaluating the corresponding 
Clebsch-Gordon coefficients leads to 
\begin{itemize}
\item
a representation of the $(B-L)$-photon as 
\begin{eqnarray}  
B_\mu=\bar u_1 \gamma_\mu u_1+\bar u_2 \gamma_\mu u_2
+\bar u_3 \gamma_\mu u_3 
\label{eqpho5}
\end{eqnarray} 
This stems from the representation 
$A_1$ on the right hand side of 
eq. (\ref{eqt1}) and from the corresponding 
Clebsch-Gordon coefficient \cite{griff}
\begin{eqnarray}  
V(T_1,T_1,A_1;i,j,1)=\frac{1}{\sqrt{3}} \delta_{ij} 
\label{ewqpho5}
\end{eqnarray} 
\item 
a representation of the gluon octet stemming from 
the remaining part $E \oplus T_1 \oplus T_2$ of 
the decomposition eq. (\ref{eqt1}). Namely, 
the CG-coefficients can be written in terms of 
the Gell-Man $\lambda$-matrices as 
\begin{eqnarray}  
V(T_1,T_1,T_1;i,j,k)&=&\frac{1}{\sqrt{6}} \epsilon_{ijk} \\
                    &=&\frac{i}{\sqrt{6}}\lambda_{7,5,2 ij} \quad for \quad k=1,2,3\\
V(T_1,T_1,T_2;i,j,k)&=&\frac{1}{\sqrt{6}} |\epsilon_{ijk}|\\
                    &=&\frac{1}{\sqrt{6}}\lambda_{6,4,1 ij} \quad for \quad k=1,2,3\\
V(T_1,T_1,E;i,j,1)&=&\frac{1}{2}\lambda_{8ij}\\
V(T_1,T_1,E;i,j,2)&=&\frac{1}{2}\lambda_{3ij}\\
\label{eqgg145}
\end{eqnarray} 
Note that the difference in the coefficients $\frac{1}{2}$ of $V(T_1,T_1,E)$ 
and $\frac{1}{\sqrt{6}}$ of $V(T_1,T_1,T_{1,2})$ is an artefact of the normalization factor 
$\sqrt{dim(D_3)}$ in eq. (\ref{eqt153}). All in all we obtain 
\begin{eqnarray}  
G_{3\mu}&=&\bar u_1 \gamma_\mu u_1-\bar u_2 \gamma_\mu u_2 \\
G_{8\mu}&=&\frac{1}{\sqrt{3}}(\bar u_1 \gamma_\mu u_1+\bar u_2 \gamma_\mu u_2-2\bar u_3 \gamma_\mu u_3) 
\label{eqgg45}
\end{eqnarray} 
and similarly for the other $\lambda$-matrices. 

The fact that formally the same bilinear combinations are created 
as needed in $SU(3)_{color}$-QCD is no accident but has 
to do with the fact that $S_4=T_d$ is a subgroup 
$T_d \subset SO(3) \subset SU(3)$. 
The result is therefore an 
elaboration on the claim formulated in \cite{lampe2} that the 
appearant tetrahedral symmetry of quarks and leptons is able 
to provide 'germs' of the Standard Model gauge interactions. 
\end{itemize}

It should further be noted that there is no problem of 
antifermions being involved here, because on the 
$S_4$ level there is no difference in the treatment 
of fermion-fermion and fermion-antifermion bilinears, 
because the group tensor product states do not care 
whether they are formed with particles or antiparticles. 

Nevertheless, one could have the suspicion of being 
cheated here in that one obtains complex fields from 
real representations of a discrete group. To be on the safe side, 
one may embed these considerations in the framework 
of the SU(4) model presented in section 4. 
In that model the physical vector bosons will 
be states in the representation 
${\bf (\bar 4 \otimes \bar 4 \otimes \bar 4\otimes \bar 4) 
\otimes (4 \otimes 4 \otimes 4 \otimes 4)}$. 
What is done in this section is to select 
the 12 physical vector bosons among the 
$4^8$ states in that representation 
by applying the exclusion principle 
('any physical particle must be a permutation state') 
proposed in section 4. 

As a next step the results eqs. (\ref{eqt1})-(\ref{eqgg45}) 
have to be projected from $S_4$ to $A_4$ of the vector bosons. 
This can be done by symmetrization in the family (u,c,t) and the 
isospin (up,down) degrees of freedom. Doing that the 
gluons turn out all right, but the $(B-L)$-photon is 
still missing its lepton contributions. 

The point is that 
$A_4$ has a 3-dimensional representation $T$ (for which 
9 symmetry functions are needed), a 2-dimensional 
representation $E$ (with only 2 functions because it is 
separably degenerate) and the totally symmetric representation $A$. 
Interpreted on this basis we 
obtain from the RHS of eq. (\ref{eqt1}): \\
i) the symmetry function for the totally symmetric representation $A$ \\
ii) the two symmetry functions for the representation $E$ \\
iii) 6 of the 9 $T$-functions (3 from $T_1$ and 3 from $T_2$). 

The 3 missing $T$-functions, which will be used to describe the weak 
bosons, can be obtained, for example, from the product 
\begin{equation} 
E\otimes E = A_1 \oplus A_2 \oplus E
\label{eqe1}
\end{equation}
Namely, taking $\mu$ and $\nu_\mu$ as basis functions for E on 
the LHS and evaluating the corresponding Clebsch-Gordon coefficients 
leads to 
\begin{itemize}
\item
a representation of the $(B-L)$-photon as 
$B_\mu=\bar \nu_\mu \gamma_\mu \nu_\mu +\bar \mu \gamma_\mu \mu$ 
which is due to the $A_1$-term in eq. (\ref{eqe1}) 
and, after symmetrization over the family index, 
gives in fact the missing lepton part of 
the quark-lepton symmetrized representation of $B_\mu$. 
\item 
a representation of the weak boson triplet stemming from 
the remaining part $A_2 \oplus E$ of the decomposition eq. (\ref{eqe1}). 
Namely, the CG-coefficients $V(E,E,A_2)$ and $V(E,E,E)$ are given by 
\begin{eqnarray}  
V(E,E,A_2;1,1,1)=0                   & & V(E,E,A_2;1,2,1)=\frac{1}{\sqrt{2}} \label{exg1456}\\
V(E,E,A_2;2,1,1)=-\frac{1}{\sqrt{2}} & & V(E,E,A_2;2,2,1)=0 
\label{exg145}
\end{eqnarray} 
and
\begin{eqnarray}  
V(E,E,E;1,1,1)=-\frac{1}{2} & & V(E,E,E;1,2,1)=0\\
V(E,E,E;2,1,1)=0            & & V(E,E,E;2,2,1)=\frac{1}{2} \\
V(E,E,E;1,1,2)=0            & & V(E,E,E;1,2,2)=\frac{1}{2}\\
V(E,E,E;2,1,2)=\frac{1}{2}  & & V(E,E,E;2,2,2)=0
\label{exxg145}
\end{eqnarray} 
leading to the combinations
\begin{eqnarray}  
W_1  &=&\bar\mu \gamma_\mu \nu_\mu +\bar\nu_\mu \gamma_\mu \mu \label{eyyg146}\\
i W_2&=&\bar\mu \gamma_\mu \nu_\mu -\bar\nu_\mu \gamma_\mu \mu \\
W_3  &=&\bar\mu \gamma_\mu \mu -\bar\nu_\mu \gamma_\mu \nu_\mu 
\label{eyyg145}
\end{eqnarray} 
Writing the CG-coefficients eqs. (\ref{exg1456})-\ref{exxg145}) in 
terms of Pauli matrices $\sigma$ 
\begin{eqnarray}  
V(E,E,A_2;i,j,1)&=&\frac{i}{\sqrt{2}}\sigma_{2ij}\\
\frac{1}{\sqrt{2}}V(E,E,E;i,j,2)&=&\frac{1}{\sqrt{2}}\sigma_{1ij}\\
\frac{1}{\sqrt{2}}V(E,E,E;i,j,1)&=&\frac{1}{\sqrt{2}}\sigma_{3ij}
\label{exxg14115}
\end{eqnarray} 
it becomes appearant that they are formally a $SU(2)_{weak}$ triplet. 
Since the $T$-representation of $A_4$ is the restriction 
of the triplet representation to $A_4$ considered as 
a subgroup of $SU(2)_{weak}$
they can be used as the set of missing symmetry functions 
for $T$. 
\end{itemize}

As before the result eq. (\ref{eyyg146})-(\ref{eyyg145}) has to be symmetrized 
in the family and the quark and lepton degrees of freedom. 


\section{Summary}


It is a remarkable observation, that quarks, leptons and gauge bosons 
can be ordered with the help of essentially the same symmetry group, 
the permutation group $S_4$.  

Starting from that paradigma we have seen, that and how 
from the 24x24=512 possible fermion-antifermion product 
states 12 are selected to describe the gauge bosons, 
and - though lacking an understanding of 
the underlying dynamics responsible for this 
selection - by inspection of Clebsch-Gordon 
coefficients we have tried to follow the path of 
how this dynamics works on the level of gauge bosons. 

Realizing that there is a connection of the $S_4$-states 
to representations of SU(4) we have found two options 
which match the phenomenological fermion and gauge boson spectrum equally 
well: either one uses a continuous inner symmetry group like SU(4) or 
Sp(4) together with an exclusion principle or one sticks 
to the discrete permutation symmetry. 

The discussion of SU(4) suggests the existence of 
a fundamental quartet of 'tetron' constituents. 
Up to this point the new symmetry can be kept 
completely independent from spacetime symmetries.
However, since it is difficult to generate the 
spin-$\frac{1}{2}$ behavior of quarks and leptons 
from 4 such constituents by conventional means, 
in ref.\cite{lampef} a somewhat different 
viewpoint will be taken, in which $S_4$ 
is not really an inner but a micro-geometric symmetry, 
where in physical space one has clouds of 4 tetronic 
constituents which surround a tetrahedral skeleton and 
tries to generate a (discrete) spin-$\frac{1}{2}$ behavior 
from that picture. 

This scenario is complicated by the fact that 
the spatial tetrahedral symmetry should in principle 
be relativistically generalized to a subgroup 
of the Lorentz group. 
In this connection one may even speculate 
whether there is a relation of the tetrons to the graviton, 
i.e. whether the underlying unknown interaction of tetrons may 
also be used to describe gravity.\\

{\bf Acknowledgement}

Table 3 is due to Pete Taylor from Warwick University. 
I owe him many fruitful e-mail discussions.

\end{document}